\documentclass{article}
\usepackage{ijcai24}

\usepackage{times}
\usepackage{soul}
\usepackage{url}
\usepackage[hidelinks]{hyperref}
\usepackage[utf8]{inputenc}
\usepackage[small]{caption}
\usepackage{graphicx}
\usepackage{amsmath}
\usepackage{amsthm}
\usepackage{booktabs}
\usepackage{algorithm}
\usepackage{algpseudocode}
\usepackage[switch]{lineno}
\usepackage{multirow}
\usepackage{arydshln} 
\usepackage{array}
\usepackage{placeins}

\usepackage{xcolor}

\definecolor{mycolor}{RGB}{255,0,0}

\title{Depending on yourself when you should: Mentoring LLM with RL agents to become the master in cybersecurity games }

\author{
Yikuan Yan$^{1}$\thanks{These authors contributed equally.}
\and
Yaolun Zhang$^{1}$\footnotemark[1] \and
Keman Huang$^{1,2}$\thanks{Corresponding author.} \\
\affiliations
$^1$School of Information, Renmin University of China, Beijing, China\\
$^2$Cybersecurity at MIT Sloan, MIT, Cambridge, Massachusetts, USA\\
\emails
\{yanyikuan, zhangyaolun5, keman\}@ruc.edu.cn
}

\begin{document}


\maketitle

\begin{abstract}
    Integrating LLM and reinforcement learning (RL) agent effectively to achieve complementary performance is critical in high stake tasks like cybersecurity operations. In this study, we introduce SecurityBot, a LLM agent  mentored by pre-trained RL agents, to support cybersecurity operations. In particularly, the LLM agent is supported with a \textit{profile} module to generated behavior guidelines, a \textit{memory} module to accumulate local experiences, a \textit{reflection} module to re-evaluate choices, and an \textit{action} module to reduce action space. Additionally, it adopts the collaboration mechanism to take suggestions from pre-trained RL agents, including a \textit{cursor} for dynamic suggestion taken, an \textit{aggregator} for multiple mentors' suggestions ranking and a \textit{caller} for proactive suggestion asking. Building on the CybORG experiment framework, our experiences show that SecurityBot demonstrates significant performance improvement compared with LLM or RL standalone, achieving the complementary performance in the cybersecurity games. 
\end{abstract}

\section{Introduction}

Cybersecurity operations involve the participation of various entities such as attackers and defenders. With the advancement of artificial intelligence (AI), autonomous cyber operation (ACO) agents have emerged as a promising solution in cybersecurity operations \cite{vyas2023automated}. These agents continually engage in adversarial learning within network environments, enhancing their strategic capabilities. The recent proliferation of large language models (LLMs) has significantly bolstered the capabilities of autonomous agents \cite{wang2023survey}. In comparison to traditional machine learning agents, LLM agents possess extensive knowledge, enabling them to handle richer and more complex information, coupled with robust contextual and reasoning abilities \cite{lin2023swiftsage,wang2023humanoid,wang2023jarvis}. They not only surpass state-of-the-art methods as novel tools \cite{xia2023universal} but also exhibit formidable interactive capabilities as assistants or agents \cite{sandoval2023lost}.

However, LLM agents lack the specific knowledge of the local environment, incur higher training costs \cite{Hu2023enabling} and can stuck in hallucinations \cite{ji2023survey,chen2023can}, while also presenting attackers with powerful weapons, making them double-edge sword for cybersecurity \cite{chen2023can,taddeo2019trusting}. Recent research attempts to frame ACO as partially observable Markov processes (POMDP), employing reinforcement learning (RL) methods to train autonomous agents \cite{standen2021cyborg,msft:cyberbattlesim}. However, without appropriate tuning methods, RL agents tend to converge to local optima, lacking robustness and generalization capabilities despite achieving favorable results \cite{palmer2023deep}. As prior studies have demonstrated that collaborations among multiple agents can enhance team performance \cite{dong2023self,ma2023eureka}, enabling the effective collaborations between LLM agents and RL agents, which can leverage the \textit{generalization} knowledge of LLMs and the \textit{specialized} knowledge of RLs in cybersecurity scenarios, can be promising to achieve complementary performance beyond that of individual agent.

Hence, we introduce the \textbf{SecurityBot}, a collaborative framework utilizing RL agents as mentors for LLM agent to support cybersecurity operations. We integrate four effective modules – \textit{profiles, memory, reflection and action} – into the LLM. Simultaneously, we propose a dynamic mechanism consisting of a \textit{cursor} to dynamically incorporate RL agents' suggestions, an \textit{aggregator} to rank suggestions from different RL agents, as well as a \textit{caller} to proactively request mentoring from RL agents. We conduct experiments on the open-source ACO research platform, CybORG \cite{standen2021cyborg}, comparing the \textit{red team (attacker) task} and \textit{blue team (defender) task} performance among: (1)independently executing RL or LLM agents (\textbf{Independent}), (2) collaboration between a LLM agent and a RL agent (\textbf{Single-Mentor}), and (3) collaboration between a LLM agent and multiple RL agents (\textbf{Multi-Mentors}). Our experimental results demonstrate that the developed \textbf{SecurityBot} can effectively improve both the red team and blue team task performance compared to independent LLM or RL approaches. Furthermore, while mentoring from multiple RL agents can be beneficial, the guidance of poorly performing RL agents may be noise to, and result into unstable performance.

\begin{itemize}
    \item We introduce \textbf{SecurityBot}, a  mechanism to enable the effective collaboration between LLM and RL agents, to leverage RL agents as mentors to accelerate learning for LLM agents and achieve complementary performance.
    \item The collaboration of LLM and RL agents demonstrates performance improvement in both red team and blue team tasks, providing a promising solution of autonomous agents for cybersecurity operations.
\end{itemize}
\section{Related Work}
\subsection{LLMs for cybersecurity operations}

Given the rapid development of LLMs and the eager to incorporate advanced AIs into cybersecurity operations \cite{iannone2022secret}, recent studies have started to explore using LLMs to enhance cybersecurity while several evidences also reveal abusing LLMs to bring advanced threats, making it a double-edged sword \cite{taddeo2019trusting,yao2023survey}

\subsubsection{LLM to enhance cybersecurity}
LLMs demonstrate advantages in both code security and data security \cite{noever2023can,ali2023huntgpt,qi2023loggpt}. For example, Fuzz4All \cite{xia2023universal} utilizes LLMs as input generators and mutation engines to generate diverse inputs for various programming languages, achieving an 36.8\% coverage improvement compared to previous state-of-the-art techniques. 

Additionally, compared to traditional machine learning approaches, LLMs possess more powerful natural language processing and contextual understanding capabilities, allowing them to elevate cybersecurity from specific to more macroscopic tasks. For example, some researches\cite{deng2023pentestgpt,pearce2023examining} utilized these capabilities in specific security tasks to enhance effectiveness, while McIntosh et al.\cite{mcintosh2023harnessing} take a further step to compared GPT-generated Governance, Risk, and Compliance (GRC) policies with those from established security vendors and government cybersecurity agencies, recommending GPT integration into companies' GRC policy development.

\subsubsection{LLMs' double-edged sword role for cybe
security}

However, applying LLMs to cybersecurity is a double-edged sword \cite{taddeo2019trusting}: being generative in nature can lead to \textit{hallucinations}—the generation of misleading or incorrect content, and can not effectively discern security-related fallacies, which can be catastrophic for high-stakes security tasks \cite{ji2023survey}. These errors can compromise sensitive operations, thereby introducing substantial risks \cite{chen2023can}. As LLMs become more integrated into security frameworks, the imperative to address and mitigate these challenges grows ever more critical.

Furthermore, LLMs present attackers with powerful weapons. Recent studies have demonstrated that LLMs can significantly enhance attacks across hardware \cite{yaman2023agent}, software and network \cite{chen2023can} levels, especially that LLMs possess human-like reasoning capabilities, making user-level attacks even more severe \cite{yao2023survey,falade2023decoding,botacin2023gpthreats}.
\subsection{Collaboration mechanisms to improve LLMs}

Recent studies have explored different mechanisms to support LLM's collaborations with others, either LLM-based or RL-based agents, including:

\subsubsection{Role-based multi-LLM-agent collaboration} Within LLM-based multi-agent systems, LLM-based agents are assigned with different roles, like decomposing complex tasks, identifying errors, and collecting multiple perspectives. Then they collaborate with each other through a series of processes to resolve complex tasks such as software developments \cite{dong2023self,qian2023communicative,hong2023metagpt}, sociological investigations \cite{park2023generative,wang2023humanoid,zhang2023exploring}, simulation of multiplayer games \cite{sandoval2023lost,xu2023exploring} and various challenges (such as logical reasoning, stock advice, blog composing, and more) \cite{li2023camel,wu2023autogen,talebirad2023multi}.  In particularly, different role-based agents exchange ideas through conversation, enforce tools to undertake tasks, garner feedback, leading to successful collaboration \cite{wang2023survey}.

\subsubsection{Dual-process-based LLM-RL collaboration} The dual process theory highlights that human cognition consists of two mental systems where System 1 is autonomous and characterized by rapid intuition, while System 2 controls slow, deliberate thinking \cite{wason1974dual,kahneman2011thinking}. Grounded on this theory, SwiftSage introduces a framework that enables a small RL model, acting as the System 1 component, to collaborate with an LLM-Based agent, acting as the System 2 component. This structure effectively solve complex problems while reducing the cost of inference \cite{lin2023swiftsage}.

\subsubsection{LLM setting guidance to support RL}
Some recent studies incorporate the LLM to generate or learn the reward function for RL agents, aiming at simplifying the reward function design process \cite{ma2023eureka,carta2022eager}. For example, \cite{DBLP:conf/iclr/MicheliAF23,DBLP:conf/iclr/KwonXBS23,du2023guiding} use LLM as a proxy reward function to guide RL agents in environments without clear reward signals. Additionally, \cite{brohan2023can,dasgupta2023collaborating} utilize the LLM-Based agent as a planner to guide RL agent in complex and dynamic environments.

\subsubsection{RL acting as expert to guide LLM's decision} 
LLM demonstrate powerful generalization abilities, but under specific scenario, they perform poorly due to the lack of expert trajectories. In contrast, RL models possess expert trajectories. Hence, \cite{Hu2023enabling,wan2022retroformer} use RL methods assist the LLM-Based agent in comprehending the environment, mastering  expert-like actions, which results in better effects and lower interaction cost instructions.
\\~

Overall, LLMs has demonstrated promising potential in enhancing cybersecurity operations while their double-edged sword role raise specific concerns. Additionally, recent studies have explored different collaborations with LLMs but it is still in its early stage, especially for cybersecurity operations. Hence, using the cybersecurity adversarial game as the research context, we design a framework with four plugin modules and three collaboration mechanisms to power LLMs for cybersecurity operations, including both acting as attackers and defenders.

\section{Cybersecurity Adversarial Game and Pre-trained RL Agents}

Before detailing our design, we start with briefly introducing our research context: the cybersecurity adversarial game. In particularly, we have constructed a cybersecurity adversarial game utilizing CybORG \cite{standen2021cyborg}, an exemplary RL-based Autonomous Cyber Operation (ACO) gym. ACO supports the creation of decision-making agents for both the blue team (defender) and the red team (attacker) in adversarial scenarios, and conveys structured and unstructured information, enabling the adaptation of both RL and LLM agents. 


\subsection{Cybersecurity Adversarial Games}

\begin{figure}[h]
    \centering

    \includegraphics[width=1\linewidth]{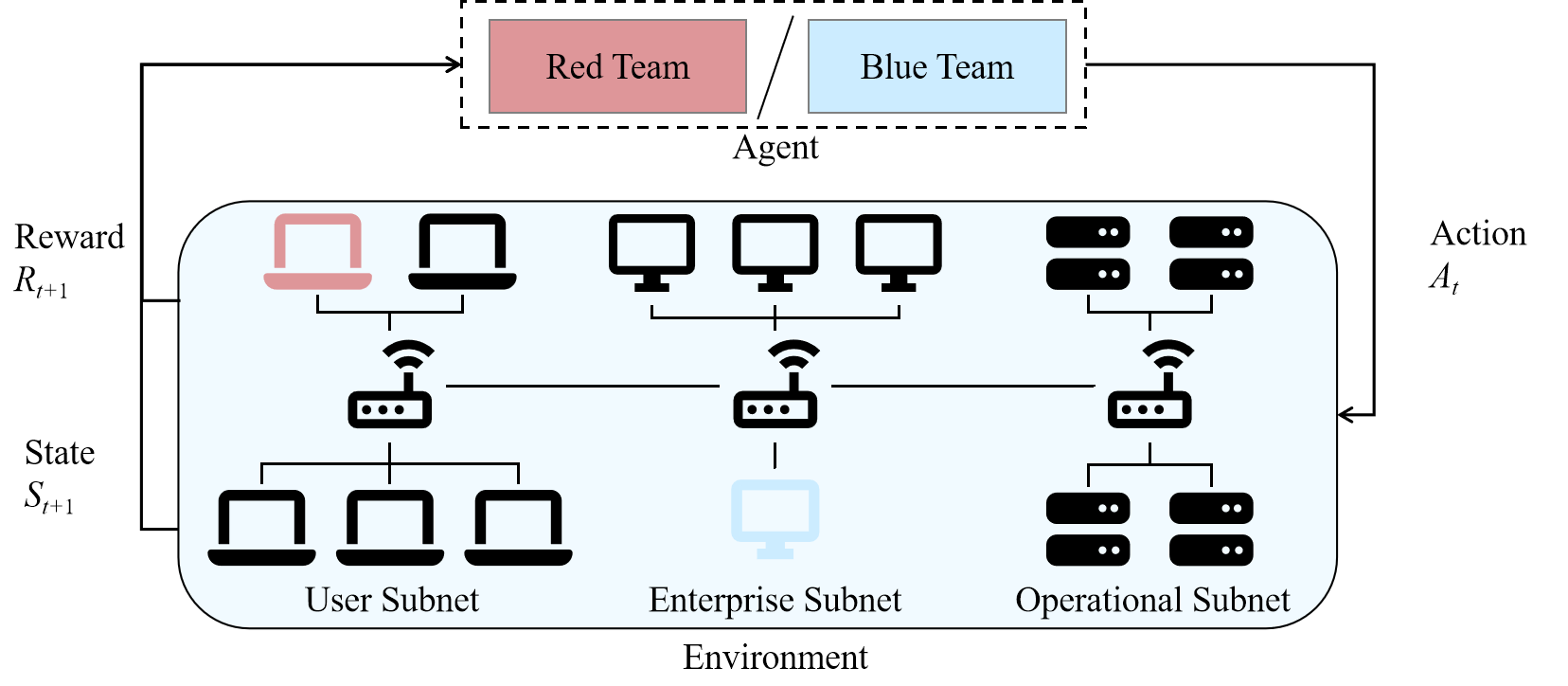}
    \caption{A POMDP cybersecurity adversial game. The red host in \textit{User Subnet} represents the foot node of the red team. The blue host in \textit{Enterprise Subnet} represents the defender host of the blue team.}
    \label{fig:3-1}
\end{figure}

The scenario adopted in this study is derived from TTCP CAGE Challenge 1 \footnote{https://github.com/cage-challenge/cage-challenge-1}, an open challenge on CybORG in 2021. As illustrated in Figure \ref{fig:3-1}, the red and blue teams compete in a simulated network environment, which can be modeled as a partially observed Markov process (POMDP). At each step, the red team and blue team take actions sequentially in the environment, causing changes in the environmental state. 


\textbf{Environment \& Observation.} The environment comprises a network consisting of 13 hosts divided into three subnets. The red team commences from the footnode in the user subnet without knowledge of any other hosts. The blue team possesses information about all hosts but lacks knowledge regarding the red team's access status to the hosts.

For both the red and blue team RL agents, their vector observation at each step encompasses: (1) whether the last action is success, (2) whether the adversary has operated on a specific host, and (3) the red team's access status of a specific host. Note that the observation is not guaranteed accurate due to the presence of an adversary.

\begin{figure}[h]
    \centering
    \includegraphics[width=1\linewidth]{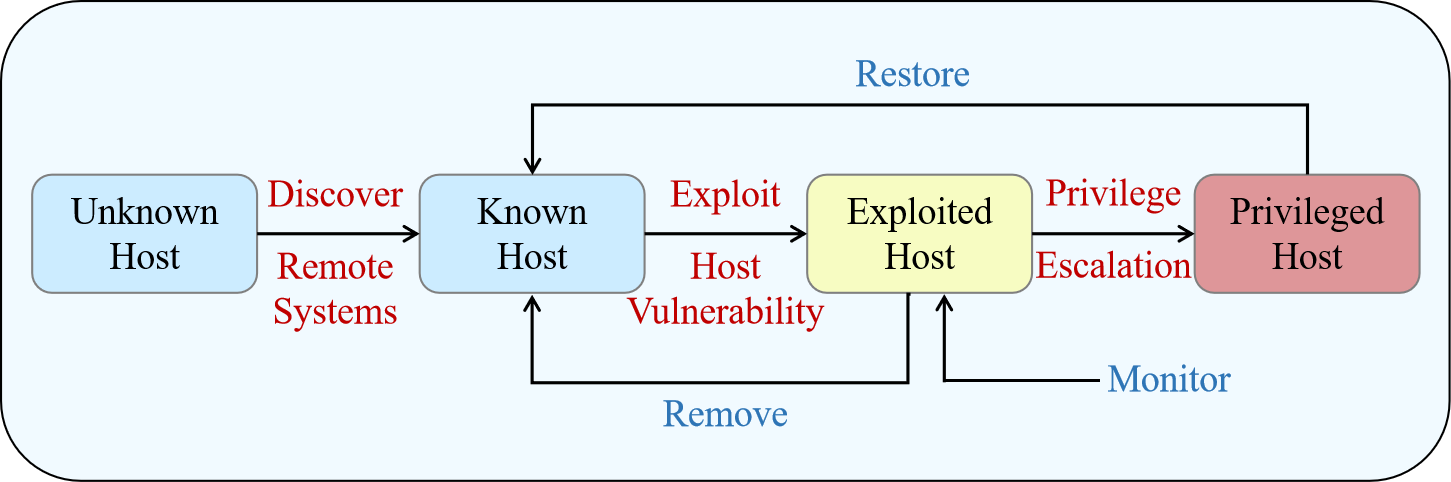}
    \caption{Action-Status Transition. Red text represents red team actions, blue text represents blue team actions.}
    \label{fig:3-2}
\end{figure}

\textbf{Action \& Reward.} As shown in Figure \ref{fig:3-2}, the two teams each have three reciprocal actions that cause transitions in the host's access status. The red team achieves lateral movement between subnets by discovering new hosts through connections from the privileged host. We set the game to be zero-sum, which means that the blue team's reward is the opposite of the red team's reward. The reward at each step is based on the extent of red team's exploitation,

\begin{equation}
    Reward_{t} = \sum_{i=1}^{n} V_{i,t}\times A_{i,t}
\end{equation}

where $V_{i, t}$ and $A_{i,t}$ represents the value and the access status of $host_{i}$ at step $t$ respectively. 

\subsection{Pre-trained RL Agents}

In this study, we choose three representative RL algorithms to train red team and blue team agents\footnote{Our framework is flexible to use other RL algorithms.}: 
\begin{itemize}
    \item \textbf{A3C} (Asynchronous Advantage Actor-Critic) \cite{mnih2016asynchronous} combines policy gradient and value function methods by asynchronously training multiple agents to improve efficiency.
    \item \textbf{DQN} (Deep Q-Network) \cite{mnih2013playing} utilizes deep neural networks to approximate the Q-value function to guide the agent's decisions.
    \item \textbf{PPO} (Proximal Policy Optimization) \cite{schulman2017proximal}, a policy gradient method, ensures stability through proximal policy optimization, restricting the magnitude of policy updates. 
\end{itemize}

The RL-based environment facilitates agent's training. Red team and blue team agents are trained separately, with one agent trained at a time. For agent's adversary, we applied the fixed-strategy agents provided in CybORG. In particular, when training a red-team RL agent, we use a blue-team agent with fixed strategy which randomly performs \textit{Remove} or \textit{Restore} operations when encountering suspicious hosts during each \textit{Monitor} action. When training a blue-team RL agent, the red-team agent as the adversary gains access to network nodes one by one based on a breadth-first strategy. Our approach aligns with the conventional RL training paradigm, wherein the agent takes an action at each step, assimilates new observations and associated rewards, and incrementally refines its strategic framework.


\section{SecurityBot: an LLM-based agent mentored by RL agents}

\begin{figure}[ht]
    \centering
    \includegraphics[width=1\linewidth]{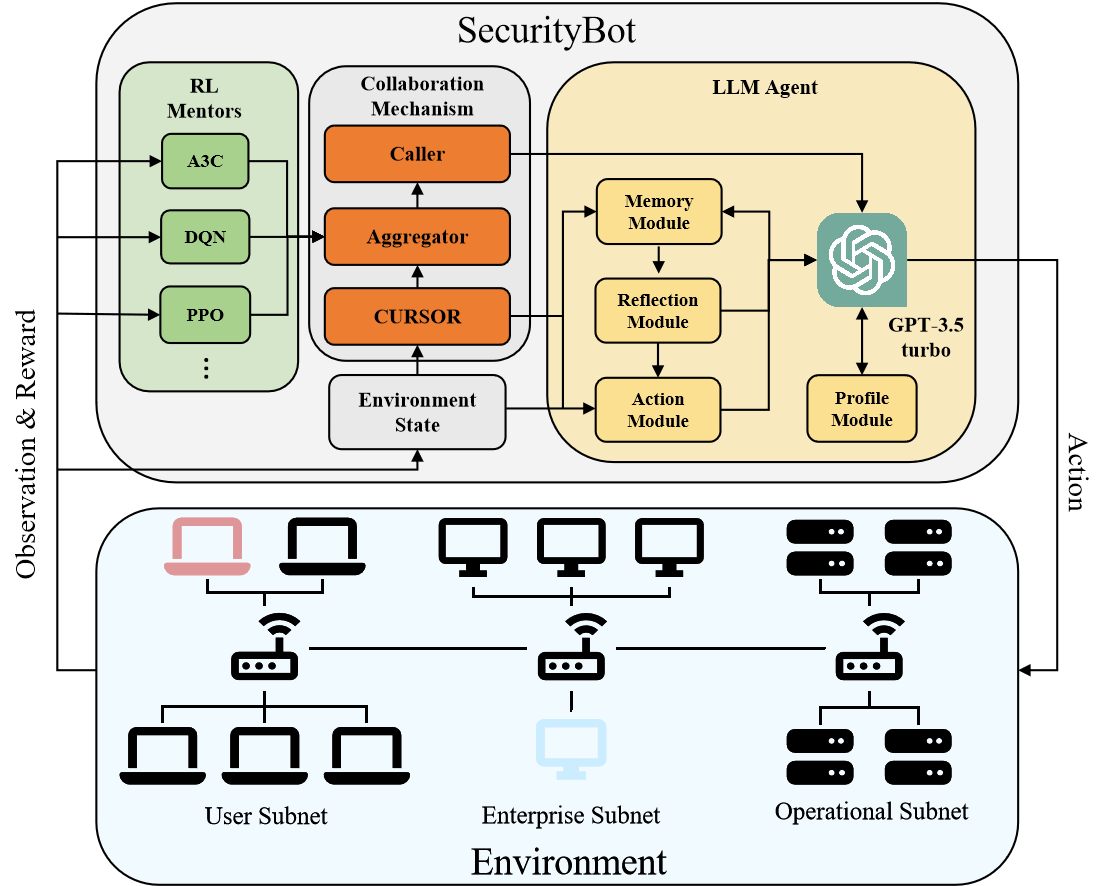}
    \caption{The Framework of SecurityBot: LLM-based RLs-mentoring Agent for Cybersecurity Operation}
    \label{fig:1}
\end{figure}

As shown in \ref{fig:1}, our SecurityBot contains three main parts: a LLM-based Agent, the pre-trained RL agent pool as mentors and their collaborative mechanisms.

\subsection{LLM Agent Design}

Building upon the LLM, GPT 3.5-turbo, our LLM agent includes four plugin modules for decision making in each step:

\subsubsection{Profile module}

\begin{figure}
    \centering
    \includegraphics[width=1\linewidth]{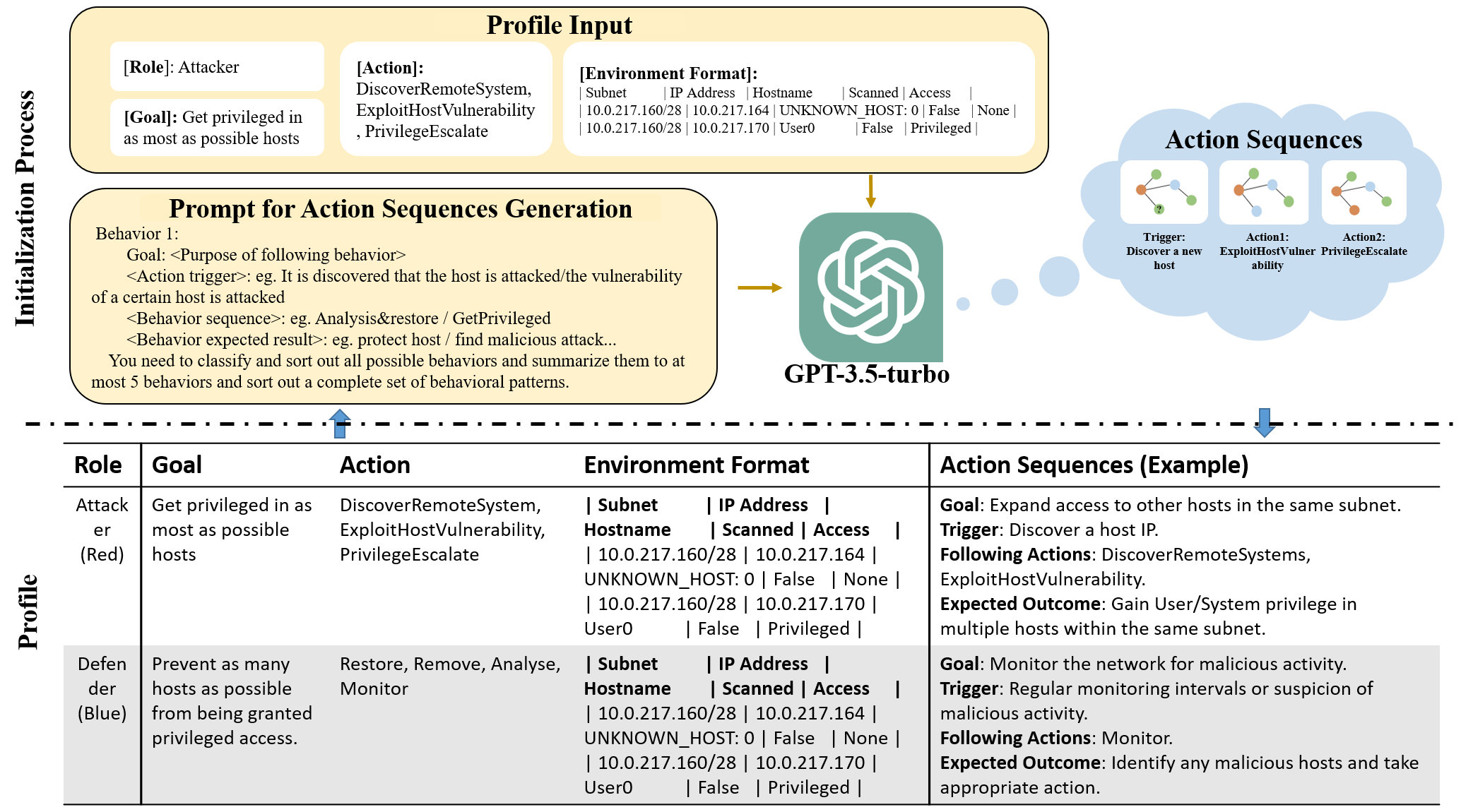}
    \caption{The illustration of profile module, including the example of roles, goals, actions, environment format and the generated behavior guidance (the bottom part) as well as the process to generate the behavior guidance (the upper part).}
    \label{fig:2}
\end{figure}

As shown in Figure \ref{fig:2}, the Profile module initializes each agent's role, goal, and available actions depending on its role. In particular, we design a prompt including the expected format for the observed environment as the input, and the expected output which is an action sequence including a series of actions with its goal, trigger, following actions, and expected outcome. When initializing the LLM agent, we use this prompt, together with the assigned goal, action, and environment format, to ask the LLM to generate an action sequence and add it to the profile, serving as the global behavior guidance for the LLM agent.

\subsubsection{Memory module}

The Memory module is used to store past experiences and search the related ones for decision making in each step. 

\textbf{\textit{Memory Storage.}} The memory module stores records including the timestamp, observed environment, action taken, and the outcome including the action status (success or failure) and its reward. In particular, when storing each memory record, the LLM agent  rates its \textbf{Importance} by prompting the LLM to score it on a scale of 0 to 10. 

\textbf{\textit{Memory Searching.}} When searching memories to support action selection in each step, the LLM agent will calculate each memory record's \textit{Relevance} and  \textit{Freshness}:

\begin{itemize}
    \item \textbf{Relevance:} measuring its environment's similarity with the current one. We transformed each environment into vectors, and then calculate their cosine similarity.
    \item \textbf{Freshness:} measuring its freshness, represented as the reciprocal of its timestamp gap with the current step.
\end{itemize}

Finally, we calculate the product of the \textit{importance}, \textit{relevance} and \textit{freshness} for each memory record and select the top two as the \textit{memory input} for LLM when making decision.

\subsubsection{Action module}

The Action module plays a crucial role in guiding the LLM agent to take valid action for each step. In particularly, given the observed environment and the available actions provided by the profile, this module will generate the action space with all the potential actions that the agent could take. 


\subsubsection{Reflection module}

Given the complex and dynamic environment, as the adversary agent may change the environment but is unobservant to the LLM agent, the LLM agent may encounter dilemmas situation, reflected as repetitive actions or diminishing rewards. For example, the red agent might persist in attacking a host in the network, even when such an action has been proven futile. Hence, the reflection module is designed to monitor the dilemma status and trigger the reflection process.

\begin{figure}
    \centering
    \includegraphics[width=1\linewidth]{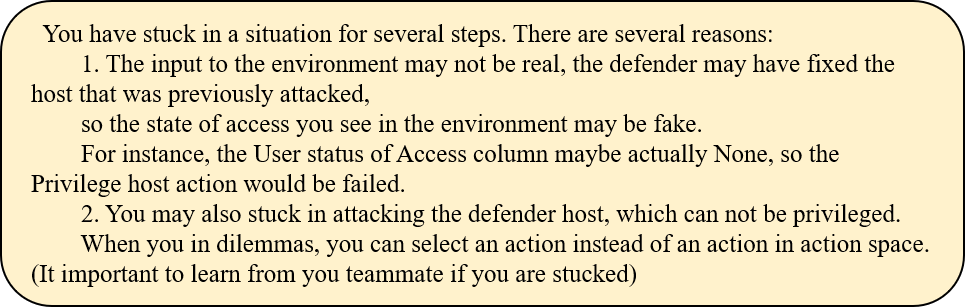}
    \caption{The prompt for Red Agent from the reflection module to motivate the LLM to choose other attack actions.}
    \label{fig:reflect}
\end{figure}

\textbf{\textit{Dilemmas Monitor.}} At every step, the Reflection module evaluates both the Reward List and the Action List from the previous steps. If there is no increase in rewards or if the agent repeats an action, the module will collect these suspicious actions, including the series of actions associated with those records, and then activate the reflection process.

\textbf{\textit{Reflection Process.}} The reflection process will pass these suspicious actions to the Action module and remove them if they are included in the generated action space. Additionally, as shown in Figure \ref{fig:reflect}, the process provides the LLM with a prompt, elucidating that the agent is stuck in the dilemmas and providing the possible reasons to guide the LLM to choose other actions to get out of the dilemma situation.

\subsection{Collaboration with RL agents}

Using RL agents as mentors to guide the LLM agent is critical for SecurityBot to achieve better performance. More specifically, as shown in Figure \ref{fig:4-2}, we design three collaboration mechanisms: 

\begin{figure}[h]
    \centering
\includegraphics[width=1\linewidth]{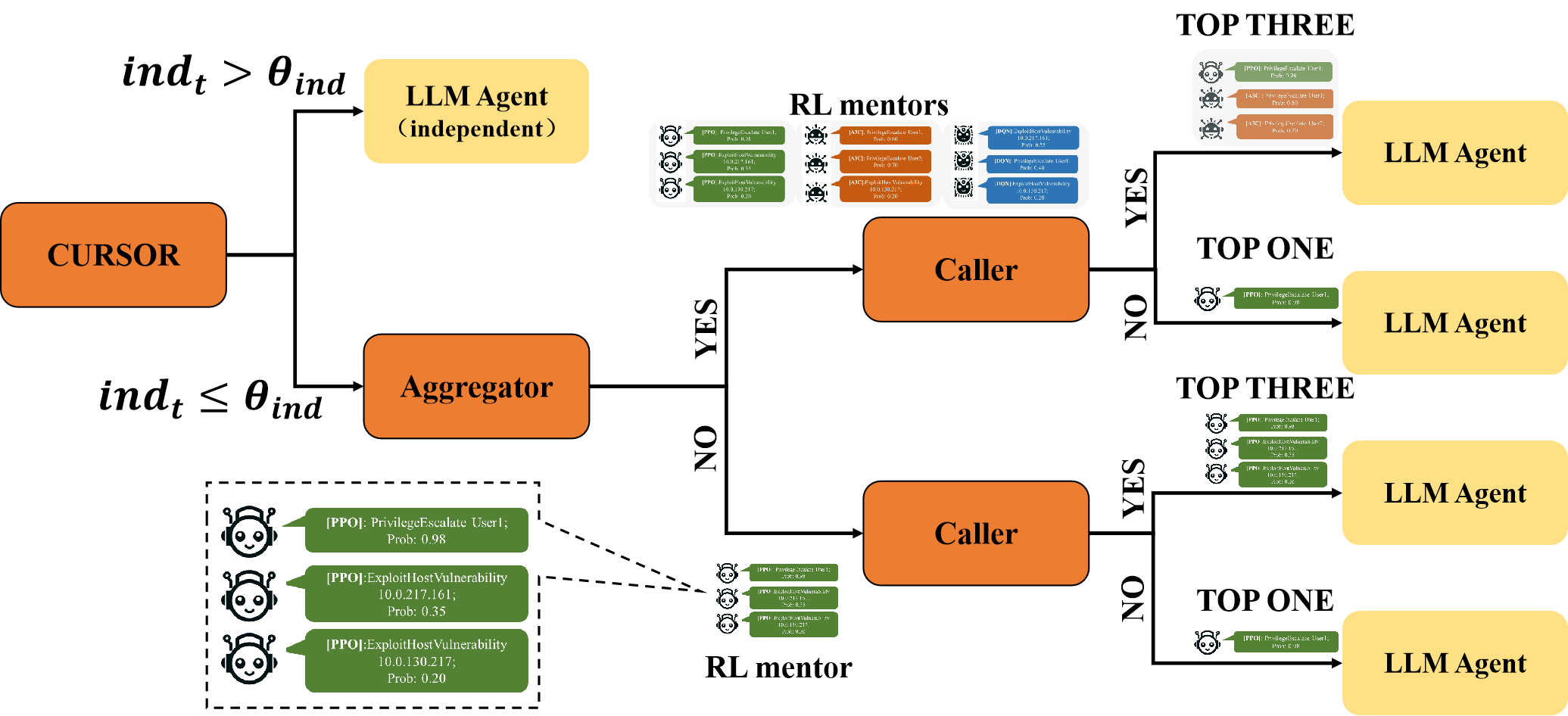}
    \caption{Mechanisms to collaboration with RL agents. Different color refers to suggestions of different RL mentors.}
    \label{fig:4-2}
\end{figure}

\subsubsection{Cursor: growing to be independent}
Firstly, the RL agents are pre-trained in the same environment particularly to guarantee that they can provide knowledge to mentor the LLM agent to make better decisions, especially in the early stage when LLM agents contain no information regarding the environment.
However, as time goes by, the LLM agent, with its capacity to understand complex environments and accumulated experience, can surpass the RL mentors (which we will report later).  Hence, we design the mechanism \textit{Cursor} to decide whether the LLM agent should take suggestions from RL agents. 

In particular, for each step $t$, the Cursor module will calculate an independence value $ind_{t}$ and only when the independence value $ind_{t}$ is below the given threshold $\theta_{ind}$, the LLM agent will consider suggestions from RL agents. Otherwise, the LLM agent will make the decision by itself. Hence, the Cursor module will adjust an independence value $ind_{t}$ in a way to reflects the tendency to rely on itself and consider the mentor's suggestion when it proves beneficial. As detailed in Equation \ref{equ:cur}, we adopt the monotonically increasing function \( f_x \) (Equation \ref{equ:fun}) so that $part_1$ reflects the trend to rely on the LLM itself. $part_2$ represents the trend of gaining reward from previous actions while $part_3$ is the signal function (Equation \ref{equ:sgn}) indicating whether the action is chosen when considering suggestions from mentors. In other words, if the LLM agent achieves an increasing reward without mentoring by the RLs, we would increase the independence value to make LLM agent more independent. Note that we introduce parameter $\alpha$ to control the change race and $\theta_{lr}$ to represent the minimal reward increment that we would expect the LLM agent to gain. 

\begin{equation}
\label{equ:cur}
\resizebox{1\hsize}{!}{$
\begin{aligned}
    ind_{t} &= ind_{t-1} + \underbrace{(f_{t}-f_{t-1})}_{\text{part\_1}} \\
    &\quad + \underbrace{\min(\alpha \times ind_{t-1},(r_{t-1}-r_{t-2}-\theta_{lr}))}_{\text{part\_2}} 
    &\quad \times \underbrace{\operatorname{sgn}(ind_{t-1}-\theta_{ind})}_{\text{part\_3}}
\end{aligned}
$}
\end{equation}

\begin{equation}
\label{equ:fun}
    \resizebox{.25\linewidth}{!}{$
            \displaystyle
           f_x = \frac{1}{1 + e^{-kx}}
        $}.
\end{equation}

\begin{equation}
\label{equ:sgn}
    \resizebox{.45\linewidth}{!}{$
            \displaystyle
           sgn (x)= \{  \begin{array}{c} 
  -1 \ \ \ if \ x>0\\ 
  1  \ \ \ \ \ otherwise
\end{array}
        $}.
\end{equation}

\subsubsection{Aggregator: ranking suggestions from multiple mentors}

Rather than relying on only one RL agent, the LLM can refer to multiple RL agents, as different RL agents may catch different aspects of the tasks. Hence, we further introduce the \textit{aggregator} mechanism to aggregate suggestions from multiple RL agents. In particular, given the top three suggestions from all the RL mentors associated with confidence, the multi-mentor mechanism will sort them based on the confidence and the top one will be presented to the LLM and the top three actions will be provided while in dilemmas. In such a case, the LLM agent does not necessarily always get suggestions from one specific RL agent during the whole task duration.

\subsubsection{Caller: asking for help proactively when in dilemma}

As discussed above, when the LLM agent encounters a dilemma, the reflection module will be activated. Furthermore, beyond activating the reflection process, the LLM agent can further refer to RL agents for support. Unlike referring to RL mentors' input in normal situation where only one suggestion is provided, we will provide the top three confident suggestions from the RL mentors.

\section{Experiments and Results}

\subsection{Experiment Setup}
\textbf{Environment.} Following the setup of Cage Challenge 1, we set the maximum number of steps in one episode, i.e., a complete round of the game, to be 100. As mentioned earlier, we set two reward parameters as shown in Table \ref{reward_parameters}: (1) \textit{Host value}. The hosts in different subnets have different values, and (2) \textit{Access state}. The higher the access state of a host, the higher the proportion of host value obtained by the red team.


\begin{table}[H]
\centering
\caption{Parameters of agent reward.}
\label{reward_parameters}
{\small
\begin{tabular}{cccc}
\toprule
\textbf{Host Subnet(V)} & \textbf{Reward} & \textbf{Access status(A)} & \textbf{Reward} \\ 
\midrule
User Subnet            & 0.1            & Unknown/Known             & 0              \\
Enterprise Subnet      & 1.0            & Exploited                 & 0.5            \\
Operational Subnet     & 10.0           & Privileged                & 0.89           \\ 
\bottomrule
\end{tabular}
}
\end{table}

\textbf{RL Training.} The RL training process is based on the \textit{Ray RLlib}, a Python library for RL\footnote{We focuses on the collaboration between RL agents and LLM agents, rather than training a better RL agent. Hence, we choose the adversary using the simplest strategy and default parameters without parameter tuning for the training algorithms are used. All specific algorithm parameters can refer to https://github.com/ray-project/ray/blob/master/rllib/algorithms/}. Each training process consists of a total of 100 iterations (4000 episodes in total). 

\textbf{LLMs Setup.} 
We leverage OpenAI's gpt-3.5-turbo API for building the LLM Agent. All the temperatures are set to 0 to restrict the format of LLM output.
\begin{itemize}
    \item Reflection. If the action is repeated in the last three steps, or if there is no increase in reward values in the last five steps, the reflection mechanism will be triggered.
    \item Cursor. $\theta_{ind}$ is set to 0.6. $\theta_{lr}$ is set to 0.3. $\alpha$ is set to 0.3. $k$ in $f(x)$ is set to 0.0135.
\end{itemize}

\textbf{Measurement.} We consider the following measurements.
\begin{itemize}
    \item Step reward. The reward of each step.
    \item Collaboration Rate ($Col$). The rate of cooperation with RL agents.
    \item Dilemma Rate ($DR$). The rate of collaborating with RL agents triggered by trapping into dilemma situation.
    \item Accept Rate ($AR$). The rate that LLM agent takes RL mentor's suggestion, indicating the extent to which LLM Agent relies on RL mentors.
    \item Accept Rate in dilemma ($AR_d$). The rate LLM agent take suggestions when trapping into dilemma, showing the ability of RL mentors to help LLM Agent out.
\end{itemize}


\textbf{Experiment Group.} We incrementally add collaboration modules and assess their performance for both the red and blue team. For each group, we run the simulation for 5 times and calculate the average.
\begin{itemize}
    \item Independent. Each RL agent (A3C, DQN, PPO) and our designed LLM agent conduct the task independently.
    \item Single-mentor. The LLM agent cooperate with a single RL agent (A3C\&LLM, DQN\&LLM, PPO\&LLM).
    \item Multi-mentors. LLM agent cooperate with all three different RL agents (MultiMentor).
\end{itemize}

\subsection{Performance in Red Team Task}

% ...


In the red team task, LLM agents and RL agents exhibited distinct action patterns, indicative of differing knowledge bases. While collaborative synergy can surpass individual agent performance, optimal collaboration is achieved when RL agents exhibit superior performance. However, when the LLM agents considers suggestions from multiple RL agents, it struggles to efficiently process this information, leading to a decline in collaborative performance.\footnote{We smoothed the data using exponential smoothing and calculated confidence intervals}

\begin{figure}[tb]
    \centering
\includegraphics[width=1\linewidth]{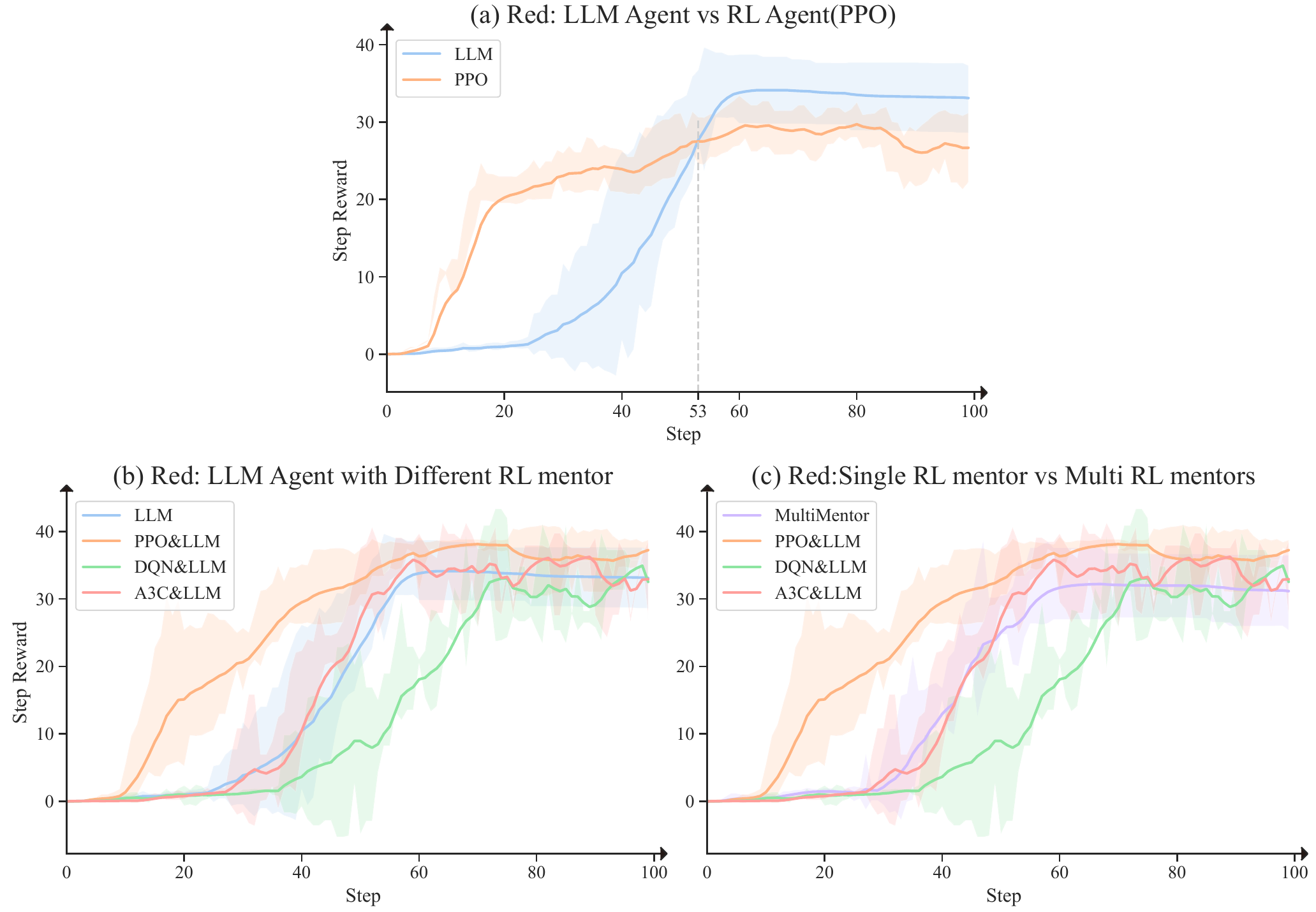}
    \caption{\textbf{Result of red team task.} (a) Comparison between LLM and PPO. They have different performances in different stages. (b)Single RL mentor result. PPO\&LLM surpasses all others. (c)Comparison between Multi and Single RL mentor. PPO\&LLM still performs best}
    \label{fig:red_result}
\end{figure}

\subsubsection{Complement knowledge of LLM agents and RL mentors}
As depicted in Figure \ref{fig:red_result}(a)\footnote{The performance of the three RL agents varies, while the PPO agent demonstrating superior performance. Due to space limitation, we only report the performance for PPO agent.}, the reward curves of the LLM agent and the PPO agent intersect: the PPO agent rapidly accumulates rewards early on, leveling off later. This behavior arises from the PPO agent gaining environmental knowledge during training, recognizing the high value of hosts in the Operational subnet. While exhibiting depth-first characteristics, insufficient training causes it to converge to a local optimum. 

Conversely, the LLM agent, despite a modest early-stage reward, achieves rapid growth, outperforming the PPO agent in the later stage. The LLM agent's behavior follows a breadth-first pattern, accumulating more exploited hosts in the network efficiently avoiding defender blocks, resulting in a higher reward.

Taking a step further, we find that LLM agents outperform PPO agents in single-step gains occurring at step 53 on average, where we differentiate the early and later stages. In later stage, we find that RL mentors always repeat one action, while LLM agent, with the reflection module, can prevent the problem. This can be the reason why RL mentors' performance is worse than LLM agent in the stage.

\subsubsection{Amplification effect of single-mentor  mechanisms}
A stronger RL mentor enhances collaborative performance, otherwise it may slows down the LLM agent's process. As shown in Figure \ref{fig:red_result}(b), PPO and A3C agents exhibit superior collaborative performance compared to LLM agents alone, and in particular, the PPO\&LLM group demonstrating a synergistic $1+1>2$ effect throughout the process, as well as getting into the rapid-growth phase much earlier. 

\begin{table}[ht]
\centering
\caption{Cooperation metric of red team task}
{\small
\begin{tabular}{c c c c}
\toprule
\multicolumn{1}{c}{Metric}     & \multicolumn{1}{c}{PPO\&LLM}     & A3C\&LLM & DQN\&LLM\\
\midrule
$ $&$Early \backslash Later$&$Early \backslash Later$&$Early \backslash Later$\\
$Col$ & $61.5\% \backslash 33.3\%$ & $78.8\% \backslash 56.2\%$ &$53.8\% \backslash 43.7\%$\\
$DR$ & $50.0\% \backslash 100.0\%$ & $34.1\% \backslash 55.6\%$& $39.3\% \backslash 80.9\%$\\
$AR$ & $50.0\% \backslash 63.6\%$ & $29.2\% \backslash 51.9\%$ & $35.7\% \backslash 57.1\%$\\
$AR_d$ & $50.0\% \backslash 63.6\%$ & $28.6\% \backslash 53.3\%$& $27.3\% \backslash 52.9\%$\\

\bottomrule
\end{tabular}
}
\label{tab:simplified_table}
\end{table}

Furthermore, the cooperation mechanism guides the LLM agent to learn from RL mentors in the early stage while seeking help in dilemmas. As shown in Table \ref{tab:simplified_table}, the LLM agent collaborates more with RL mentors in the early stages than later, satisfying our design goal. $DR$ are all higher in the later stage, meaning most collaborations with the RL agent are triggered by the dilemmas situation. Interestingly, $AR$ and $AR_d$ values are both higher in the later stage, meaning in the later stage, despite outperforming the RL mentor, the LLM agent relies more on the RL mentor's suggestions if needed. 

\subsubsection{Noise from multi-mentors}

We explored whether the LLM agent could gain more knowledge from recommendations of multi-mentors. In our setup, assistance from multiple RL mentors is not necessary helpful. As shown in Figure \ref{fig:red_result}(c), while the performance of multi-tutors slightly outperforms LLM alone, it falls short of the LLM\&PPO group. We observed that 75.61\% of suggestions from RL mentors originated from DQN, but only 5.41\% were accepted. In contrast, 34.61\% of PPO's suggestions were accepted. Moreover, only 15.85\% of all RL suggestions were accepted, markedly lower than the acceptance rate in a single mentor scenario. This disparity illuminates the high confidence suggestion from the low performance mentor became a \textit{noise} for the LLM agent.

\subsection{Performance in Blue Team Task}

\begin{figure}[tb]
    \centering
    \includegraphics[width=1\linewidth]{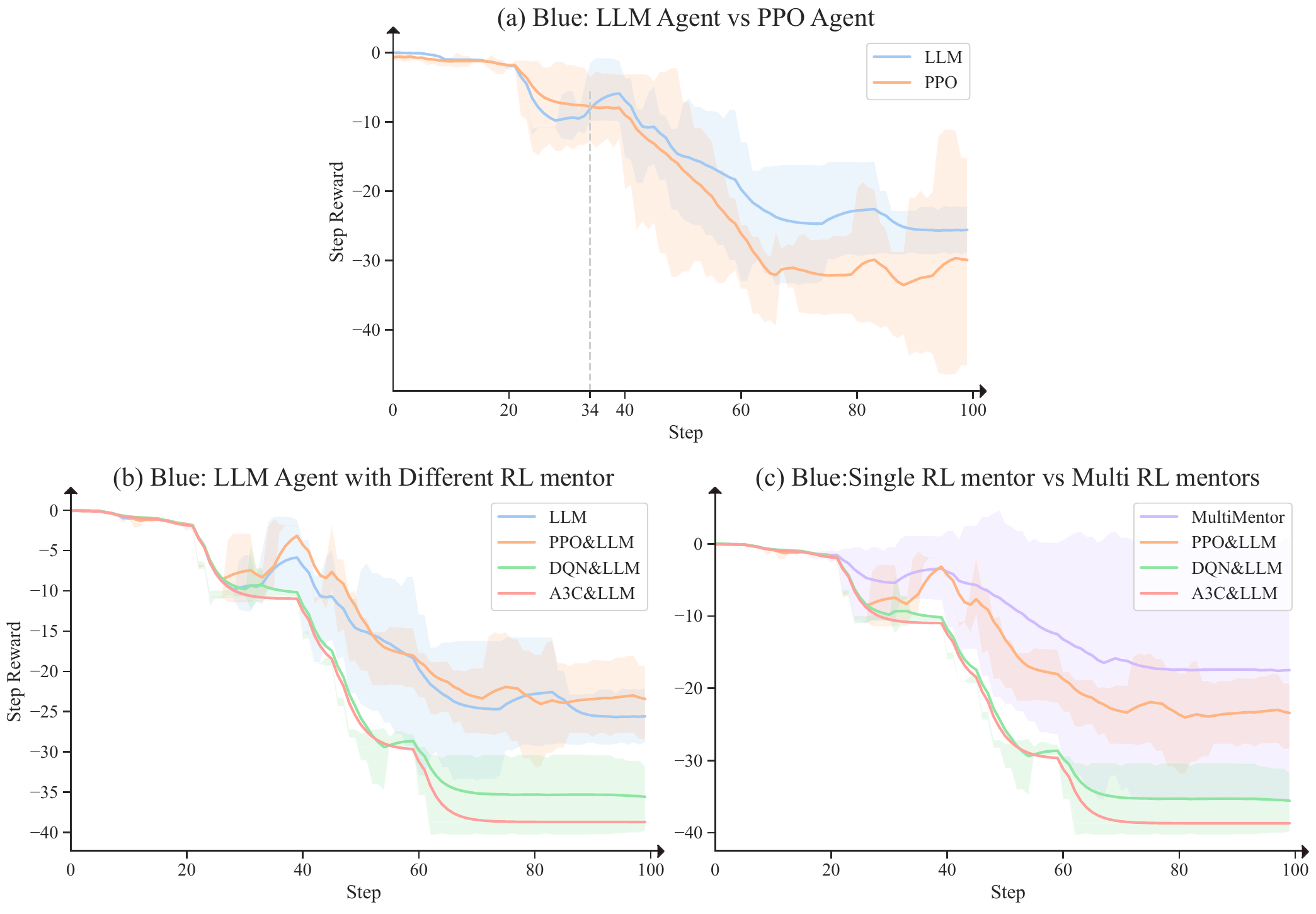}
    \caption{\textbf{Result of blue team task.} (a)Comparison between LLM and PPO. LLM outperform PPO in Blue Team Task. (b) Single RL mentor result. PPO\&LLM perform slightly better than LLM. (c) Comparison between Multi and Single RL mentor.Multi-mentor perform best on average, but not stable enough.}
    \label{fig:5-3}
\end{figure}

\subsubsection{A helpful but narrower complementary knowledge}

As shown in Figure \ref{fig:5-3} (a), the LLM agents demonstrate performance similar to the PPO agent during the early stages. But after a brief period of divergence, the LLM agent consistently outperforms the PPO agent. We observe the similar situation in the case of single mentor. As shown in Figure \ref{fig:5-3} (b), although PPO\&LLM group demonstrates a marginally superior performance over LLM agent, this advantage is not observed in other groups. These results indicate a narrower knowledge gap between LLM and RL agents in blue team task, may due to the fact that the whole network environment is used for pre-training RL agents and  provided to LLM agent.

Additionally, as reported in Table \ref{tab:blue team task}, the LLM agent would accept RL mentors' suggestions in the early stages. While in the later stage, both A3C\&LLM and DQN\&LLM groups show little interest in RL mentor's suggestion except trapped in dilemmas. Conversely, we can observe consistently higher $AR$ rates in the later stages for PPO\&LLM. This discrepancy indicates the LLM agent's capability in identifying the suggestion quality and the importance of providing high quality suggestion to improve the LLM agent's effectiveness.

\begin{table}[h]
\caption{Cooperation result in blue team task}
\centering
{\small
\begin{tabular}{cccc}
\toprule
\multicolumn{1}{c}{Metric}     & \multicolumn{1}{c}{PPO\&LLM}     & A3C\&LLM & DQN\&LLM\\
\midrule
$ $&$Early \backslash Later$&$Early \backslash Later$&$Early \backslash Later$\\
$Col$ & $48.1\% \backslash 22.9\%$& $78.8\% \backslash 33.3\%$ & $71.2\% \backslash 16.7\%$ \\
$DR$ & $40.0\% \backslash 45.5\%$& $43.9\% \backslash 100.0\%$& $27.0\% \backslash 100.0\%$\\
$AR$ & $100.0\% \backslash 81.9\%$ & $53.7\% \backslash 28.6\%$ & $91.9\% \backslash 28.6\%$\\
$AR_d$ & $100.0\% \backslash 60.0\%$ & $31.3\% \backslash 28.6\%$& $100.0\% \backslash 28.6\%$ \\

\bottomrule
\end{tabular}
}
\label{tab:blue team task}
\end{table}

\subsubsection{Outstanding but unstable performance of multi-mentors}
In contrast to the red team task, as shown in Figure \ref{fig:5-3} (c), the incorporation of multiple RL mentors enhances the average performance of the blue team task beyond that of both the LLM agents and the PPO\&LLM group. However, this configuration exhibits  instability demonstrated as a larger confidence intervals. While it effectively defends nearly all hosts at times, in some instances, its performance is comparable to that of a single LLM. Notably, the LLM Agent accepts less than 5\% of suggestions from RL mentors, predominantly originating from DQN. One reason behind this is that the most confident RL suggestions are not consistently the most effective, especially when provided by multiple mentors. 

Additionally, in the blue team task, LLM agents showcase a superior understanding of the environment, often acting independently in most situations. Particularly in scenarios where the LLM agent successfully defends almost all hosts, it appears to disregard unreliable suggestions from multiple RL mentors, opting to make critical decisions autonomously.

\section{Conclusion and Future Work}

This study presents SecurityBot, a LLM agent powered by mentoring from pre-trained RL agents for cybersecurity operations. In particular, with the designed plugin modules, including the profile, memory, reflection and action modules to enhance the LLM, and three collaboration mechanisms, including a cursor, an aggregator and a caller, to effectively collaborate with pre-trained RL agents, the LLM agent achieve significant performance improvement in both cyber attack and defense tasks. Although RL agents can learn local knowledge effectively through pre-training, the LLM agent  can surpass them through learning the environments in the later stage. This confirms that our designed LLM agent can be a promising solution to support cybersecurity operations.

While RL agents' suggestions can be helpful, especially when the LLM agent trapped in dilemmas, as observed in our result, weak RL agents may serve as a noise to distract the LLM agent. Further research can design advanced aggregating strategies to extract the essence and discard the dross from RL agents. Furthermore, while we aim at empowering LLM with plugin modules and collaborating it with RL agents, we did not fintune the LLM or optimize the RL agents. Future studies can fintune a better LLM model specific for cybersecurity operations and train optimized RL agents, which could further improve the SecurityBot's performance.

\appendix





\bibliographystyle{named}

\end{document}